
\magnification=\magstep1
\def\figindents{\leftskip=3 true pc \rightskip=2 true pc}
\def\verline#1#2#3{\rlap{\kern#1mm\raise#2mm
                   \hbox{\vrule height #3mm depth 0 pt}}}
\def\mirrorline#1#2#3{\rlap{\kern#1mm\raise#2mm
                   \hbox{\vrule height #3mm width 0.1 pt depth 0 pt}}}
\def\horline#1#2#3{\rlap{\kern#1mm\raise#2mm
                   \vbox{\hrule width #3mm depth 0 pt}}}
\def\Verline#1#2#3{\rlap{\kern#1mm\raise#2mm
                   \hbox{\vrule height #3mm width 0.7pt depth 0 pt}}}
\def\Horline#1#2#3{\rlap{\kern#1mm\raise#2mm
                   \vbox{\hrule height 0.7pt width #3mm depth 0 pt}}}
\def\putbox#1#2#3{\setbox117=\hbox{#3}
                  \dimen121=#1mm
                  \dimen122=#2mm
                  \dimen123=\wd117
                  \dimen124=\ht117
                  \divide\dimen123 by -2
                  \divide\dimen124 by -2
                  \advance\dimen121 by \dimen123
                  \advance\dimen122 by \dimen124
                  \rlap{\kern\dimen121\raise\dimen122\hbox{#3}}}
\def\leftputbox#1#2#3{\setbox117=\hbox{#3}
                  \dimen122=#2mm
                  \dimen124=\ht117
                  \divide\dimen124 by -2
                  \advance\dimen122 by \dimen124
                  \rlap{\kern#1mm\raise\dimen122\hbox{#3}}}
\def\topputbox#1#2#3{\setbox117=\hbox{#3}
                  \dimen121=#1mm
                  \dimen123=\wd117
                  \divide\dimen123 by -2
                  \advance\dimen121 by \dimen123
                  \rlap{\kern\dimen121\raise#2mm\hbox{#3}}}
\def\plot#1#2{
  \putbox{#1}{#2}{$\scriptstyle.$}}
\def\point#1#2{
  \putbox{#1}{#2}{\bf.}}
\def\sqr#1#2{{\vcenter{\vbox{\hrule height.#2pt
            \hbox{\vrule width.#2pt height#1pt \kern#1pt
                  \vrule width.#2pt}\hrule height.#2pt}}}}
\def\square
 {\mathop{\mathchoice{\sqr{12}{15}}{\sqr{9}{12}}{\sqr{6.3}{9}}{\sqr{4.5}{9}}}}
\overfullrule=0pt
\def\w{$\cal W$}
\def\wt{\widetilde}
\def\q#1{\lbrack #1\rbrack}

\def\smno{\smallskip\noindent}
\def\meno{\medskip\noindent}
\def\bigno{\bigskip\noindent}
\def\pano{\par\noindent}

\def\o#1{\overline{#1}}

\def\sw{${\cal SW}$}
\def\pt{\partial}

\def\sst{\scriptstyle}
\def\cl{\centerline}

\def\BNT{\,\hbox{\hbox to -0.2pt{\vrule height 6.5pt width .2pt\hss}\rm N}}
\def\BRT{\,\hbox{\hbox to -0.2pt{\vrule height 6.5pt width .2pt\hss}\rm R}}
\def\BZT{{\rm Z{\hbox to 3pt{\hss\rm Z}}}}
\def\BZS{{\hbox{\sevenrm Z{\hbox to 2.3pt{\hss\sevenrm Z}}}}}
\def\BZSS{{\hbox{\fiverm Z{\hbox to 1.8pt{\hss\fiverm Z}}}}}
\def\BZ{{\mathchoice{\BZT}{\BZT}{\BZS}{\BZSS}}}
\def\BQT{\,\hbox{\hbox to -2.8pt{\vrule height 6.5pt width .2pt\hss}\rm Q}}
\def\BQS{\,\hbox{\hbox to -2.1pt{\vrule height 4.5pt width .2pt\hss}$
    \scriptstyle\rm Q$}}
\def\BQSS{\,\hbox{\hbox to -1.8pt{\vrule height 3pt width
    .2pt\hss}$\scriptscriptstyle \rm Q$}}

\def\BCT{\,\hbox{\hbox to -3pt{\vrule height 6.5pt width .2pt\hss}\rm C}}
\def\BCS{\,\hbox{\hbox to -2.2pt{\vrule height 4.5pt width .2pt\hss}$
    \scriptstyle\rm C$}}
\def\BCSS{\,\hbox{\hbox to -2pt{\vrule height 3.3pt width
    .2pt\hss}$\scriptscriptstyle \rm C$}}

\def\section#1{\leftline{\bf #1}\vskip-7pt\line{\hrulefill}}
\def\bibitem#1{\parindent=8mm\item{\hbox to 6 mm{\q{#1}\hfill}}}
\def\berg{1}
\def\banks{2}
\def\selfa{3}
\def\selfb{4}
\def\selfc{5}
\def\cande{6}
\def\candz{7}
\def\ciz{8}
\def\egu{9}
\def\fonti{10}
\def\fuchse{11}
\def\fuchsz{12}
\def\gep{13}
\def\gepq{14}
\def\gre{15}
\def\int{16}
\def\klsch{17}
\def\kreuzm{18}
\def\kreuza{19}
\def\kreuzs{20}
\def\kreuzl{21}
\def\lerche{22}
\def\lut{23}
\def\lyne{24}
\def\lynz{25}
\def\lynd{26}
\def\mart{27}
\def\qiu{28}
\def\sche{29}
\def\schz{30}
\def\scht{31}
\def\schwarz{32}
\def\schwim{33}
\def\vw{34}
\def\zog{35}

\font\Large=cmr12 scaled \magstep3

\rm
\nopagenumbers
\pano
{\rightline {\vbox{\hbox{hep-th/9503129}
                   \hbox{BONN-TH-95-05}
                   \hbox{IFP-504-UNC}
                   \hbox{March 1995} }}}
\vskip 1.5cm
\centerline{\Large A Note On ADE String Compactifications }
\vskip 2cm
\centerline{{Ralph Blumenhagen${}^1$}\ \  and \ \
          {Andreas Wi{\ss}kirchen${}^2$}}
\vskip 0.5cm
\centerline{${}^1$ \it Institute of Field Physics, Department of Physics
and Astronomy,}
\centerline{\it University of North Carolina, Chapel Hill NC 27599-3255, USA}
\vskip 0.1cm
\centerline{${}^2$ \it Physikalisches Institut der Universit\"at Bonn,
Nu{\ss}allee 12, 53115 Bonn, Germany}
\vskip 2cm
\centerline{\bf Abstract}
\meno
We address the question whether so-called $m$-invariants of the $N=2$ super
Virasoro algebra can be used for the construction of reasonable
four-dimensional string models. It turns out that an infinite subset of those
are pathological in the sense that -- although $N=2$ supersymmetric -- the
Ramond sector is not isomorphic to the Neveu-Schwarz sector. Consequently,
these two properties are independent and only requiring both guarantees an
$N=1$ space-time supersymmetric string spectrum. However, the remaining 529
consistent spectra -- 210 of them are mirrors of Gepner models and 76 real
orbifolds -- show exact mirror symmetry and are contained in a recent
classification of orbifolds of Gepner models.
\footnote{}
{\pano
${}^1$ e-mail:\ blumenha@physics.unc.edu
\pano
${}^2$ e-mail:\ wisskirc@avzw01.physik.uni-bonn.de}
\vfill
\eject
\footline{\hss\tenrm\folio\hss}
\pageno=1
\section{1.\ Introduction}
\pano
In 1988, Gepner [\gep] initiated the construction of four-dimensional
$N=1$ space-time supersymmetric string vacua with an extended gauge group
$E_6\times E_8$ using explicitly $N=2$ supersymmetric conformal field
theories (SCFT) in the internal sector [\lut,\lyne,\lynz]. In the following
years his approach of tensoring unitary models of the $N=2$ super Virasoro
algebra ($VIR_{N=2}$) adding up to $c=9$ has been extended to orbifolds
[\fonti,\fuchse,\fuchsz,\zog] and simple current constructions [\schz,\scht].
Another, field theoretic, approach started with $N=2$ supersymmetric
Landau-Ginzburg models [\mart,\vw] which is more related
to the geometric description in terms of Calabi-Yau manifolds [\cande].
Recently, a classification of all orbifolds including discrete torsion of
so-called ADE invariants of the $VIR_{N=2}$ has been completed in
[\kreuza,\kreuzs]. It turned out that an earlier stochastic search using
the simple current technique [\schz,\scht] was almost exhaustive.
\pano
It is known that the ADE invariants  are not a complete set of modular
invariant partition functions of the $VIR_{N=2}$ [\ciz,\gepq,\qiu].
There also exist the so-called $m$-invariants. In this note we investigate
which subset of string vacua can be obtained by these further invariants.
It is clear that all these models can also be obtained by an orbifold or
simple current construction [\gepq]. However, there appear some interesting
features. First, although $N=2$ supersymmetric, a huge set of these
$m$-invariants do not yield reasonable $N=1$ space-time supersymmetric string
models. The reason is that these models have an extended \w-symmetry, the
spectral flow automorphism of which does not any longer map the Neveu-Schwarz
$(NS)$ sector (space-time bosons) onto the Ramond $(R)$ sector (space-time
fermions). Thus, this example {\sl explicitly} shows that $N=2$ supersymmetry
and the existence of the gravitino vertex operator are independent conditions
and only both of them are sufficient for $N=1$ space-time supersymmetry
[\lerche]. The necessity of those conditions has been derived in 1988 by
Banks et al.\ [\banks]. The second feature is that these string models built
up from $m$-invariants exhibit an exact mirror symmetry, which is given by a
simple exchange of $m$-invariants. Since all 529 consistent string spectra we
found are contained in the classification of [\kreuza,\schz], our calculation
can be regarded as an independent check.
\meno
\section{2.\ Modular invariants of the $\bf VIR_{N=2}$}
\pano
In this section we briefly review some facts about the unitary series
of the $VIR_{N=2}$ [\gep,\gepq,\qiu]. By realizing  $VIR_{N=2}$ as a
product of parafermions times a $U(1)$ current
$$  VIR_{N=2}={SU(2)_k\over U(1)}\times U(1),\eqno(2.1) $$
one obtains the following grid of unitary representations:
$$\eqalignno{
& c={3k\over k+2},\quad
  h^l_{m,s}={l(l+2)-m^2\over 4(k+2)}+{s^2\over 8}\ \ {\rm mod}\ 1,\quad
  q^l_{m,s}={m\over k+2}-{s\over 2}\ \ {\rm mod}\ 2, &(2.2)\cr
 & k\in\BNT,\
 l\in\lbrace 0,\dots,k\rbrace,\ m\in\lbrace -k-1,\dots,k+2\rbrace,\
   s\in\lbrace -1,\dots,2\rbrace,\ l+m+s=0\ {\rm mod}\ 2.&\cr }$$
To get a non-redundant set of primary fields $\Phi^l_{m,s}$ one has to take
into account the reflection symmetry
$$    \Phi^l_{m,s}=\Phi^{k-l}_{m+k+2,s+2},\quad m=m\ {\rm mod}\ 2(k+2),
      \quad s=s\ {\rm mod}\ 4.\eqno(2.3)$$
Then, the supersymmetric characters are given by
$$      \chi^l_m(z,\tau,u)=\chi^l_{m,s}(z,\tau,u)\pm
          \chi^l_{m,s+2}(z,\tau,u),\quad s=0\ {\rm for}\
            NS,\widetilde{NS}\,{\rm and}\,
         s=-1\ {\rm for}\ R,\widetilde{R} \eqno(2.4)$$
where the minus sign has to be chosen for the $\widetilde{NS}$ and
$\widetilde{R}$ sector. The corresponding superprimary fields are denoted
by $\Phi^l_{m}$. Under modular transformations
$$     S:(z,\tau,u)\to ({z\over\tau},-{1\over \tau},u+{z^2\over 2\tau}),
        \quad T:(z,\tau,u)\to (z,\tau+1,u) \eqno(2.5)$$
these characters behave in a very simple way:
$$\eqalignno{
 \chi^l_{m,s}({z\over\tau},-{1\over \tau},u+{z^2\over 2\tau})&=
             K\sum_{l',m',s'}{\rm sin}\left(\pi{(l+1)(l'+1)\over k+2}\right)
             \,e^{\pi i mm'\over k+2}\,e^{-\pi i ss'\over2}\,
                \chi^l_{m,s}(z,\tau,u), &\cr
            \chi^l_{m,s}(z,\tau+1,u)&=e^{2\pi i \left(h^l_{m,s}-
        {c\over 24}\right)}\,\chi^l_{m,s}(z,\tau,u)&(2.6)\cr} $$
with some constant $K$ only depending on $k$,
i.e.\ the level $l$ transforms like an $SU(2)_k$ character, $m$ as a
$\Theta_{m,k+2}$ and $s$ as a $\Theta_{s,4}$ function. Thus, a huge set of
modular invariant partition functions is given by the product ansatz
$$    Z=\sum_{l,l',m,m',s,s'}\, N_{l,l'}\, M_{m,m'}\, P_{s,s'}\,
            \chi^l_{m,s}\o{\chi}^{l'}_{m',s'} \eqno(2.7)$$
with $N,M,P$ representing modular invariant combinations of the three
corresponding models. The matrices $ N_{l,l'}$ are classified in an
ADE scheme [\ciz] and the possible  $m$-invariants $M_{m,m'}$ of the system of
$\Theta_{\cdot,k}$-functions are labeled by divisors of $k$ [\gepq].
For any factorization $k=\alpha\cdot\beta$ there exists a modular invariant
partition function
$$     M_{m,m'}={1\over 2}\sum_{x\in\BZ_{2\beta}, y\in\BZ_{2\alpha}}
         \delta_{m,\alpha x+\beta y} \delta_{m',\alpha x-\beta y}.\eqno(2.8)$$
All these are simple current invariants due to Schellekens and Yankielowicz
[\sche]. The simple current is the primary field corresponding to
$\Theta_{2\beta,k}$ and the length of an orbit turns out to be $\alpha$.
Thus, except the three cases $E_6,E_7,E_8$ all modular invariants
(2.7) can be obtained by modding out a simple current and therefore due to
[\kreuzs] by an orbifold construction.
\meno
\section{3.\ Off-diagonal invariants for k=4j-2}
\pano
For $k=4j$ the corresponding $D_{2j+2}$ invariant of $SU(2)_k$  yields an
off-diagonal partition function of $VIR_{N=2}$ which can be interpreted as a
diagonal partition function of an extended superconformal algebra. In this
case it is a \sw$(1,j^0)$, i.e.\ the extension of $VIR_{N=2}$ by a primary
superfield of superconformal dimension $H=j$ and $U(1)$ charge $Q=0$. For
$j\in\lbrace 2,3\rbrace$ these super \w-algebras have been explicitly
constructed in [\selfa,\selfb]. However, there are also consistent solutions
for \sw$(1,j^{0})$ with $j\in\BZ+{1\over2}$. These are implied by
$m$-invariants with $k=4j-2$ and the factorization $k+2=4j=2\cdot(2j)$:
$$    Z=\sum_{ \sst m=2\atop\sst m=0\, {\rm mod}\, 2}^{4j}
|\Theta_{m,4j}+\Theta_{m-4j,4j}|^2.\eqno(3.1)$$
The corresponding partition function of $VIR_{N=2}$ is
$$\eqalignno{
             Z=&Z_{NS}+Z_{\wt{NS}}+Z_{R},&\cr
             Z_{NS}=&
              {1\over2}\sum_{\sst l=0\atop\sst l=0\,{\rm mod}\,2}^{2j-2}
               \sum_{\sst m=-l\atop\sst m=0\,{\rm mod}\,2}^{l}
                |\chi^l_m+\chi^{4j-l-2}_m|^2+
              {1\over2}\sum_{\sst l=2j\atop\sst l=0\,{\rm mod}\,2}^{4j-2}
                \sum_{\sst m=4j-l\atop\sst m=0\,{\rm mod}\,2}^{l}
                |\chi^l_m + \chi^l_{m-4j}|^2,  &\cr
              Z_{\wt{NS}}=&
               {1\over2}\sum_{\sst l=0\atop\sst l=0\,{\rm mod}\,2}^{2j-2}
                 \sum_{\sst m=-l\atop\sst m=0\,{\rm mod}\,2}^{l}
                  |\wt\chi^l_m-\wt\chi^{4j-l-2}_m|^2+
               {1\over2}\sum_{\sst l=2j\atop\sst l=0\,{\rm mod}\,2}^{4j-2}
                 \sum_{\sst m=4j-l\atop\sst m=0\, {\rm mod}\, 2}^{l}
                  |\wt{\chi}^l_m+\wt{\chi}^l_{m-4j}|^2,  &\cr
              Z_{R}=&
               {1\over2}\sum_{\sst l=1\atop\sst l=1\,{\rm mod}\,2}^{2j-3}
                \sum_{\sst m=-l+1\atop\sst m=0\, {\rm mod}\, 2}^{l+1}
                 |\chi^l_m+\chi^{4j-l-2}_m|^2+
               {1\over 2}\sum_{\sst l=2j+1\atop\sst l=1\,{\rm mod}\,2}^{4j-3}
                \sum_{\sst m=4j-l+1\atop\sst m=0\,{\rm mod}\,2}^{l+1}
                 |{\chi}^l_m+{\chi}^l_{m-4j}|^2 &\cr
              &+\sum_{\sst m=-2j+2\atop\sst m=0\,{\rm mod}\,2}^{2j}
                 |{\chi}^{2j-1}_m|^2.  &(3.2)\cr}$$
The $\wt{R}$ sector is left invariant by the modular group:
$$   Z_{\wt{R}}={1\over2}\sum_{\sst l=1\atop\sst l=1\,{\rm mod}\,2}^{2j-3}
                \sum_{\sst m=-l+1\atop\sst m=0\,{\rm mod}\,2}^{l+1}
                 |\wt\chi^l_m-\wt\chi^{4j-l-2}_m|^2+
             {1\over2}\sum_{\sst l=2j+1\atop\sst l=1\,{\rm mod}\,2}^{4j-3}
                \sum_{\sst m=4j-l+1\atop\sst m=0\,{\rm mod}\,2}^{l+1}
                 |\wt{\chi}^l_m+\wt{\chi}^l_{m-4j}|^2. \eqno(3.3) $$
The intriguing feature is that, since in the $R$ sector odd values of $m$ are
projected out, the spectral flow operator $\Phi^0_1$ does not occur. This
can be understood in the following way:\ The above partition function is the
diagonal one for the extension of $VIR_{N=2}$ by the primary superfield
$\Phi^{4j-2}_0$ of dimension $(H,Q)=(j-{1\over2},0)$, cf.\ [3] for the
notation:
$$ \Phi(Z)=\phi^0_{j-{1\over 2} }(z)+{1\over\sqrt{2}}\Big(
   \theta^+{\psi}^{-1}_{j }(z)- \theta^-{\psi}^{+1}_{j }(z)\Big)
   +\theta^+\theta^-\chi^0_{j+{1\over 2} }(z).\eqno(3.4)$$
A spin structure $\left({\scriptstyle.}\square\limits_.
\big\vert\ldots\big\vert{\scriptstyle.}\square\limits_.\right)$
on the torus is defined by different boundary conditions of the involved
fermions ($A$:\ antiperiodic, $P$:\ periodic). Besides the supercurrents
$G^\pm(z)$ there are two further fermionic fields in the symmetry algebra,
namely $\phi(z)$ and $\chi(z)$. In the following the first box represents the
boundary conditions along the two cycles of $G^\pm(z)$ and the second one
those of $\phi(z)$ and $\chi(z)$. Under modular transformations one obtains
the well-known chain
$$\left({\scriptstyle{A}}\square_A\Big\vert{\scriptstyle{A}}\square_A\right)
  {\buildrel T\over\longrightarrow}
  \left({\scriptstyle{P}}\square_A\Big\vert{\scriptstyle{P}}\square_A\right)
  {\buildrel S\over\longrightarrow}
  \left({\scriptstyle{A}}\square_P\Big\vert{\scriptstyle{A}}\square_P\right)
  .\eqno(3.5)$$
The sector $\left({\scriptstyle{P}}\square\limits_P
 \big\vert{\scriptstyle{P}}\square\limits_P\right)$
is invariant under the action of the modular group. In the $R$ sector of
the above partition function all fermions are periodic around a circle of
constant time or carry integer modes, equivalently. However, the spectral
flow automorphism [\schwim] has to be extended to the entire \w-algebra.
Denoting the action of the spectral flow by $(\cdot)',$ the only way to
preserve the primarity of the further generators is
$$\eqalignno{(L_n)'&=L_n+\eta J_n+{c\over 6}\eta^2\delta_{n,0},\quad
(G_r^\pm)'=G^\pm_{r\pm\eta},\quad(J_n)'=J_n+{c\over 3}\eta\delta_{n,0},&\cr
(F_n)'&=F_{n+Q(F)\eta}\quad\ {\rm for}\ \ F=\phi,\psi^\pm,\chi.&(3.6)\cr}$$
\pano
Since the fermionic generators $\phi, \chi$ have even $U(1)$ charge and
the bosonic ones $\psi^\pm(z)$ odd, the spectral flow
${\cal O}_{\eta={1\over 2}}$ does not any longer connect the $NS$ and the
$R$ sector. But it does generate new twisted sectors, where also the bosonic
fields carry half-integer modes. For instance, applying the spectral flow
to the $NS$ sector yields
$$\left({\scriptstyle{A}}\square_A\Big\vert{\scriptstyle{A}}\square_A\right)
   {\buildrel{\cal O}_{1\over 2}\over\longrightarrow}
  \left({\scriptstyle{A}}\square_P\Big\vert{\scriptstyle{A}}\square_A\right)
   .\eqno(3.7)$$
Now, the question arises, whether it is possible to extend the partition
function to a flow invariant one without loosing modular invariance. To
this end, one has to sum over the entire orbit generated by successive
application of the modular transformations $T,S$ and the spectral flow
${\cal O}_{1\over 2}$. The result is a sum over all 16 possible spin
structures. However, this is nothing but the diagonal partition function
for the non-extended $VIR_{N=2}$. Schematically, one has to take into
account the relation
$$ \sum_{x,y\in\lbrace{\rm P,A}\rbrace}
    \left({\scriptstyle{m}}\square_n\Big\vert{\scriptstyle{x}}\square_y\right)
     ={\scriptstyle{m}}\square_n \eqno(3.8)$$
where the box on the r.h.s.\ has to be understood as a spin structure of
the non-extended $VIR_{N=2}$, i.e.\ of the two supercurrents $G^\pm(z)$ only.
Summarizing, the model (3.2) gives an example of an $N=2$ supersymmetric CFT
in which the $R$ sector is not isomorphic to the $NS$ sector. However,
although the spectral flow ${\cal O}_{1\over 2}$ is not an automorphism
of the  model (3.2), the square of it, ${\cal O}_1$, is. It is realized by
the primary field $\Phi^0_2\ {\buildrel(2.3)\over = }\ \Phi^k_{-k}$.
Thus, if one tensors such models adding up to $c=9$, e.g.\ $(k=2)^{\otimes6}$,
and chooses the product of the partition functions of the factors, the latter
flow is still present in the theory. It can be realized in the well-known way
by the $U(1)$ current
$$ X^\pm(z)=\sqrt{6} :e^{\pm i \sqrt{3}\varphi(z)}: \quad
       {\rm with}\quad j(z)=\sqrt{3}i\pt\varphi(z).\eqno(3.9) $$
As usual, this simple current can be used to project the internal $N=2$ SCFT
onto integer $U(1)$ charges in the $NS$ and half-integer charges in the $R$
sector, respectively [\selfc,\egu,\gep]. One obtains an $N=2$ SCFT with
$c=9$ and (half-)integer $U(1)$ charges which fails to contain the internal,
holomorphic part of the gravitino vertex operator
$$\Sigma^\pm(z)= :e^{\pm i{\sqrt{3}\over 2}\varphi(z)}:.\eqno(3.10)$$
Consequently, after combining the space-time sector and the internal sector,
the next GSO projection onto odd $U(1)$ charges cannot be carried out.
In order to get a space-time supersymmetric spectrum one would have to form
supersymmetric orbits by successive application of the operator
${\cal S}_{\alpha}\, \Sigma^+ $, where ${\cal S}_{\alpha}$ denotes the
space-time spinor. Thus, this example shows that the existence of $N=1$
space-time supersymmetry is not guaranteed by $N=2$ world-sheet supersymmetry
and integer $U(1)$ charges alone. Independently, the existence of the spectral
flow operator with $\eta={1\over 2}$ in the spectrum has to be required. Then,
Gepner's construction inherently implies that these two conditions ensure
space-time supersymmetry.
\pano
Since these somehow pathological models still provide one with an $N=2$ SCFT
with $c=9$ and (half-)integer $U(1)$ charges, the question arises, whether
they can be interpreted as some nonlinear $\sigma$ model partition functions
of not Calabi-Yau type.
\meno
\section{4.\ String spectra and the mirror map}
\pano
Of course, not in all $m$-invariants the spectral flow is projected out.
For all 168 combinations of unitary models adding up to $c=9$ [\lynz] with
different ADE invariants we have used all possible  $m$-invariants in all
factors for the construction of a GSO projected, space-time supersymmetric
string compactification. However, the following two conditions put severe
constraints on the allowed combinations of invariants:
\smno
\item{$(a)$} The spectral flow operator
     $(H,Q;\o{H},\o{Q})=({3\over8},{3\over2};{3\over8},{3\over2})$
     must be contained in the $R$ sector of the $c=9$ internal SCFT.
\pano
\item{$(b)$}
     The field theoretic analogues of the $(3,0)$ and $(0,3)$ form on the
     Calabi-Yau manifold have to survive the GSO projection. These are
     exactly the (anti-)holomorphic, chiral fields with
     $(H,Q;\o{H},\o{Q})=({3\over 2},3;0,0)$ and
     $(H,Q;\o{H},\o{Q})=(0,0;{3\over 2},3)$ in the $c=9$ internal SCFT which
     correspond to spectral flow with $\eta=1$.
\smno
One can show that the condition (b) is equivalent to the requirement that the
state $(H,Q;\o{H},\o{Q})=({3\over 2},3;{3\over 2},3)$ survives the projection.
Since applying two times the spectral flow with $\eta={1\over 2}$ gives the
flow with $\eta=1$, (b) follows from (a). However, the model (3.2) shows that
both conditions are not equivalent meaning that there is not necessarily a
kind of `squareroot' of the flow with $\eta=1$. Furthermore, unlike the orbits
generating generations and antigenerations the vacuum orbit is invariant under
a $U(1)$ flip, i.e.\ $Q\to -Q$. Therefore, the fields with
$(H,Q;\o{H},\o{Q})=({3\over 2},-3;0,0)$ and
$(H,Q;\o{H},\o{Q})=(0,0;{3\over 2},-3)$ are automatically contained in the
spectrum.
\pano
Surprisingly, besides the ordinary Gepner models and their mirrors which
merely contain $m$-invariants $k_i+2=1\cdot(k_i+2)$ and $k_i+2=(k_i+2)\cdot1$,
respectively, only 76 further spectra are consistent. They are listed
in Table 1 which can be found in the appendix including the used combination
of invariants and the massless string spectrum.
The relation
$$M^{k+2=\alpha\cdot\beta}_{m,m'}=M^{k+2=\beta\cdot\alpha}_{m,-m'}\eqno(4.1)$$
for the invariants of the $\Theta$ functions (2.8) implies that all consistent
models occur in mirror pairs. To get the mirror partner one only has to
replace the $k_i+2=\alpha_i\cdot\beta_i$ $m$-invariant by the
$k_i+2=\beta_i\cdot\alpha_i$ invariant in each factor. This couples
the left and right sector in such a way that a state  $(H_L,Q_L;H_R,Q_R)$
is substituted by $(H_L,Q_L;H_R,-Q_R)$. Especially, the mirror partner
of an ordinary Gepner model can be obtained by the $k_i+2=(k_i+2)\cdot1$
invariants. Note, that this simple map yields the mirror partition function
without explicitly carrying out an orbifold construction [\berg,\gre,\lynd].
There are further interesting properties:
\smno
\item{(i)}
All Euler numbers of the 76 orbifold spectra are divisible by 12.
\pano
\item{(ii)}
There are nontrivial mirrors, i.e.\ one obtains the mirror spectrum by an
$m$-invariant different from the flip (4.1) in all factors. This phenomenon
occurs at almost all tensor products. In a few cases one needs no flip (4.1)
in any factor at all.
\pano
\item{(iii)}
Some $m$-invariants produce the same spectrum as different $l$-invariants.
This generalizes an observation made in [\lynz]. It does not happen as often
as (ii).
\smno
All 529 (i.e.\ also including the ordinary Gepner models and their mirrors)
resulting from combining $l$- and $m$-invariants are plotted in Figure 1 in
the usual manner. The dots `$\scriptstyle.$' denote Gepner spectra and their
mirrors, whereas `{\bf.}' stand for the spectra also listed in Table 1.
The plot shows exact mirror symmetry, in distinction to the results in
more general constructions [\candz,\klsch,\kreuzm,\kreuzl].
\bigno
\pano
$$\verline{100}{0}{105}
  \verline{0}{0}{105}
  \topputbox{100.1}{102.7}{$\uparrow$}
  \putbox{108.5}{105}{$n_{27}+n_{\overline{27}}$}
  \Horline{98.5}{10}{3}
  \Horline{98.5}{20}{3}
  \Horline{98.5}{30}{3}
  \Horline{98.5}{40}{3}
  \Horline{98.5}{50}{3}
  \Horline{98.5}{60}{3}
  \Horline{98.5}{70}{3}
  \Horline{98.5}{80}{3}
  \Horline{98.5}{90}{3}
  \Horline{98.5}{100}{3}
  \putbox{105}{10}{$50$}
  \putbox{105}{20}{$100$}
  \putbox{105}{30}{$150$}
  \putbox{105}{40}{$200$}
  \putbox{105}{50}{$250$}
  \putbox{105}{60}{$300$}
  \putbox{105}{70}{$350$}
  \putbox{105}{80}{$400$}
  \putbox{105}{90}{$450$}
  \putbox{105}{100}{$500$}
  \horline{0}{0}{100}
  \horline{0}{105}{100}
  \leftputbox{97}{-0.1}{$\rightarrow$}
  \putbox{100}{-2}{$\chi$}
  \Verline{10}{-1.5}{3}
  \Verline{20}{-1.5}{3}
  \Verline{30}{-1.5}{3}
  \Verline{40}{-1.5}{3}
  \Verline{50}{-1.5}{3}
  \Verline{60}{-1.5}{3}
  \Verline{70}{-1.5}{3}
  \Verline{80}{-1.5}{3}
  \Verline{90}{-1.5}{3}
  \putbox{10}{-5}{$-800$}
  \putbox{20}{-5}{$-600$}
  \putbox{30}{-5}{$-400$}
  \putbox{40}{-5}{$-200$}
  \putbox{50}{-5}{$0$}
  \putbox{60}{-5}{$200$}
  \putbox{70}{-5}{$400$}
  \putbox{80}{-5}{$600$}
  \putbox{90}{-5}{$800$}
  \plot{41.7}{16.8}
  \plot{41.1}{18.}
  \plot{44.9}{10.8}
  \plot{44.1}{12.4}
  \plot{42.9}{14.8}
  \plot{42.5}{15.6}
  \plot{41.7}{17.2}
  \plot{40.1}{20.4}
  \plot{39.9}{20.8}
  \plot{35.7}{29.2}
  \plot{35.3}{30.}
  \plot{44.7}{11.6}
  \plot{44.5}{12.}
  \plot{44.1}{12.8}
  \plot{42.9}{15.2}
  \plot{41.7}{17.6}
  \plot{39.7}{21.6}
  \plot{39.3}{22.4}
  \plot{38.1}{24.8}
  \plot{37.5}{26.}
  \plot{23.1}{54.8}
  \plot{45.3}{10.8}
  \plot{44.7}{12.}
  \plot{43.5}{14.4}
  \plot{42.9}{15.6}
  \plot{40.5}{20.4}
  \plot{39.9}{21.6}
  \plot{33.9}{33.6}
  \plot{26.1}{49.2}
  \plot{46.5}{8.8}
  \plot{42.9}{16.}
  \plot{41.1}{19.6}
  \plot{39.3}{23.2}
  \plot{35.7}{30.4}
  \plot{29.7}{42.4}
  \plot{46.5}{9.2}
  \plot{45.3}{11.6}
  \plot{44.1}{14.}
  \plot{42.9}{16.4}
  \plot{41.7}{18.8}
  \plot{40.5}{21.2}
  \plot{34.5}{33.2}
  \plot{32.1}{38.}
  \plot{44.7}{13.2}
  \plot{44.1}{14.4}
  \plot{41.7}{19.2}
  \plot{39.3}{24.}
  \plot{38.7}{25.2}
  \plot{32.7}{37.2}
  \point{47.1}{8.8}
  \plot{45.3}{12.4}
  \plot{44.7}{13.6}
  \plot{44.5}{14.}
  \point{44.1}{14.8}
  \plot{42.9}{17.2}
  \plot{41.7}{19.6}
  \plot{40.5}{22.}
  \plot{38.1}{26.8}
  \plot{36.5}{30.}
  \plot{35.7}{31.6}
  \plot{23.7}{55.6}
  \plot{47.4}{8.6}
  \plot{46.5}{10.4}
  \plot{44.1}{15.2}
  \plot{42.9}{17.6}
  \plot{42.3}{18.8}
  \plot{40.5}{22.4}
  \plot{39.3}{24.8}
  \plot{34.5}{34.4}
  \point{48.3}{7.2}
  \plot{47.7}{8.4}
  \plot{47.1}{9.6}
  \plot{45.3}{13.2}
  \point{44.7}{14.4}
  \plot{44.1}{15.6}
  \plot{42.9}{18.}
  \plot{39.9}{24.}
  \plot{39.3}{25.2}
  \plot{38.1}{27.6}
  \plot{35.7}{32.4}
  \point{47.1}{10.}
  \plot{46.5}{11.2}
  \plot{44.1}{16.}
  \plot{40.5}{23.2}
  \plot{33.3}{37.6}
  \plot{29.7}{44.8}
  \point{48.9}{6.8}
  \plot{47.7}{9.2}
  \point{46.5}{11.6}
  \plot{46.1}{12.4}
  \plot{45.9}{12.8}
  \plot{45.3}{14.}
  \plot{42.9}{18.8}
  \plot{42.1}{20.4}
  \point{41.1}{22.4}
  \plot{40.5}{23.6}
  \plot{38.1}{28.4}
  \plot{35.7}{33.2}
  \plot{28.5}{47.6}
  \plot{2.1}{100.4}
  \point{48.3}{8.4}
  \plot{46.5}{12.}
  \plot{41.7}{21.6}
  \plot{38.7}{27.6}
  \point{48.9}{7.6}
  \plot{47.7}{10.}
  \plot{46.5}{12.4}
  \plot{45.3}{14.8}
  \plot{44.1}{17.2}
  \plot{42.9}{19.6}
  \point{41.7}{22.}
  \plot{40.5}{24.4}
  \plot{39.3}{26.8}
  \plot{35.7}{34.}
  \plot{46.5}{12.8}
  \plot{41.7}{22.4}
  \plot{27.3}{51.2}
  \point{48.9}{8.4}
  \plot{47.7}{10.8}
  \plot{45.3}{15.6}
  \plot{36.9}{32.4}
  \plot{33.3}{39.6}
  \plot{46.5}{13.6}
  \plot{44.1}{18.4}
  \plot{41.7}{23.2}
  \plot{40.5}{25.6}
  \point{49.5}{8.}
  \plot{48.9}{9.2}
  \point{48.3}{10.4}
  \plot{47.7}{11.6}
  \plot{45.3}{16.4}
  \plot{44.1}{18.8}
  \plot{41.7}{23.6}
  \plot{36.3}{34.4}
  \plot{34.5}{38.}
  \plot{14.1}{78.8}
  \point{48.3}{10.8}
  \plot{47.7}{12.}
  \point{42.9}{21.6}
  \plot{41.7}{24.}
  \plot{39.3}{28.8}
  \plot{29.7}{48.}
  \point{50.1}{7.6}
  \plot{49.3}{9.2}
  \plot{47.7}{12.4}
  \plot{45.3}{17.2}
  \plot{42.9}{22.}
  \plot{48.9}{10.4}
  \plot{46.5}{15.2}
  \plot{44.7}{18.8}
  \plot{41.7}{24.8}
  \plot{34.5}{39.2}
  \plot{50.1}{8.4}
  \plot{48.9}{10.8}
  \plot{40.5}{27.6}
  \point{47.7}{13.6}
  \plot{46.5}{16.}
  \plot{45.3}{18.4}
  \plot{44.1}{20.8}
  \plot{39.3}{30.4}
  \point{51.3}{6.8}
  \point{50.7}{8.}
  \plot{50.1}{9.2}
  \point{49.5}{10.4}
  \plot{49.3}{10.8}
  \plot{48.9}{11.6}
  \plot{47.7}{14.}
  \plot{46.5}{16.4}
  \plot{45.3}{18.8}
  \plot{38.1}{33.2}
  \plot{26.1}{57.2}
  \plot{18.9}{71.6}
  \point{49.5}{10.8}
  \plot{48.9}{12.}
  \point{46.5}{16.8}
  \plot{44.1}{21.6}
  \plot{32.1}{45.6}
  \point{51.3}{7.6}
  \point{50.1}{10.}
  \plot{48.9}{12.4}
  \plot{44.1}{22.}
  \plot{42.9}{24.4}
  \plot{44.1}{22.4}
  \plot{42.9}{24.8}
  \plot{36.9}{36.8}
  \point{51.9}{7.2}
  \point{51.3}{8.4}
  \plot{50.9}{9.2}
  \plot{50.1}{10.8}
  \point{48.9}{13.2}
  \plot{46.9}{17.2}
  \point{45.3}{20.4}
  \point{49.5}{12.4}
  \plot{47.7}{16.}
  \plot{36.9}{37.6}
  \plot{51.3}{9.2}
  \point{50.7}{10.4}
  \plot{50.1}{11.6}
  \plot{48.9}{14.}
  \plot{47.7}{16.4}
  \point{46.5}{18.8}
  \plot{30.9}{50.}
  \point{51.9}{8.4}
  \point{50.7}{10.8}
  \point{47.1}{18.}
  \plot{45.3}{21.6}
  \plot{43.5}{25.2}
  \plot{50.9}{10.8}
  \plot{50.1}{12.4}
  \plot{47.7}{17.2}
  \plot{44.1}{24.4}
  \plot{42.9}{26.8}
  \plot{51.3}{10.4}
  \point{49.5}{14.}
  \point{48.9}{15.2}
  \plot{52.5}{8.4}
  \plot{51.3}{10.8}
  \plot{46.5}{20.4}
  \plot{42.9}{27.6}
  \plot{39.3}{34.8}
  \point{50.7}{12.4}
  \plot{48.9}{16.}
  \plot{47.7}{18.4}
  \plot{41.7}{30.4}
  \plot{34.5}{44.8}
  \plot{52.8}{8.6}
  \plot{52.5}{9.2}
  \point{51.9}{10.4}
  \plot{51.3}{11.6}
  \plot{50.1}{14.}
  \point{48.9}{16.4}
  \plot{47.7}{18.8}
  \point{35.7}{42.8}
  \point{51.9}{10.8}
  \plot{51.3}{12.}
  \plot{48.9}{16.8}
  \plot{39.3}{36.}
  \point{53.1}{8.8}
  \plot{52.5}{10.}
  \plot{51.3}{12.4}
  \plot{48.9}{17.2}
  \point{47.7}{19.6}
  \plot{45.3}{24.4}
  \point{50.7}{14.}
  \point{49.5}{16.4}
  \point{47.7}{20.}
  \plot{46.5}{22.4}
  \plot{53.1}{9.6}
  \plot{52.5}{10.8}
  \point{51.3}{13.2}
  \plot{50.1}{15.6}
  \plot{46.1}{23.6}
  \plot{45.3}{25.2}
  \plot{44.1}{27.6}
  \plot{53.7}{8.8}
  \point{53.1}{10.}
  \plot{46.5}{23.2}
  \plot{53.7}{9.2}
  \plot{52.5}{11.6}
  \plot{51.3}{14.}
  \plot{50.1}{16.4}
  \plot{48.9}{18.8}
  \plot{46.5}{23.6}
  \plot{52.5}{12.}
  \plot{39.3}{38.4}
  \plot{52.5}{12.4}
  \plot{50.1}{17.2}
  \plot{47.7}{22.}
  \point{45.3}{26.8}
  \plot{53.7}{10.4}
  \point{51.3}{15.2}
  \point{50.7}{16.4}
  \plot{48.3}{21.2}
  \plot{46.5}{24.8}
  \plot{50.1}{18.}
  \plot{45.3}{27.6}
  \plot{53.7}{11.2}
  \point{52.5}{13.6}
  \plot{51.3}{16.}
  \plot{44.1}{30.4}
  \point{53.7}{11.6}
  \plot{52.5}{14.}
  \point{51.3}{16.4}
  \plot{50.1}{18.8}
  \plot{26.1}{66.8}
  \plot{53.7}{12.}
  \plot{51.3}{16.8}
  \plot{36.9}{45.6}
  \plot{53.7}{12.4}
  \plot{51.3}{17.2}
  \plot{50.1}{19.6}
  \plot{47.7}{24.4}
  \plot{53.7}{12.8}
  \plot{41.7}{36.8}
  \plot{54.9}{10.8}
  \plot{54.1}{12.4}
  \point{48.9}{22.8}
  \plot{53.7}{13.6}
  \plot{52.5}{16.}
  \plot{55.3}{10.8}
  \plot{54.9}{11.6}
  \plot{54.3}{12.8}
  \plot{52.5}{16.4}
  \plot{51.3}{18.8}
  \plot{50.1}{21.2}
  \plot{46.5}{28.4}
  \point{45.3}{30.8}
  \plot{38.1}{45.2}
  \plot{54.9}{12.4}
  \plot{52.5}{17.2}
  \plot{50.1}{22.}
  \point{49.5}{23.2}
  \plot{55.5}{11.6}
  \plot{53.7}{15.2}
  \plot{55.5}{12.}
  \plot{54.9}{13.2}
  \plot{47.7}{27.6}
  \plot{55.7}{12.}
  \plot{53.7}{16.}
  \plot{52.5}{18.4}
  \plot{54.9}{14.}
  \plot{53.7}{16.4}
  \plot{53.3}{17.2}
  \plot{52.5}{18.8}
  \plot{50.1}{23.6}
  \plot{47.7}{28.4}
  \plot{42.9}{38.}
  \plot{55.5}{13.2}
  \point{53.7}{16.8}
  \point{53.1}{18.}
  \plot{56.1}{12.4}
  \plot{55.5}{13.6}
  \plot{54.9}{14.8}
  \point{52.5}{19.6}
  \point{50.7}{23.2}
  \plot{56.1}{12.8}
  \point{52.5}{20.}
  \plot{51.9}{21.2}
  \plot{55.7}{14.}
  \point{55.5}{14.4}
  \plot{54.9}{15.6}
  \point{51.3}{22.8}
  \point{50.1}{25.2}
  \plot{56.1}{14.}
  \plot{54.9}{16.4}
  \point{53.7}{18.8}
  \plot{44.1}{38.}
  \plot{56.1}{14.4}
  \plot{39.3}{48.}
  \point{56.1}{14.8}
  \plot{54.9}{17.2}
  \plot{52.5}{22.}
  \plot{56.1}{15.2}
  \plot{48.9}{29.6}
  \plot{44.1}{39.2}
  \plot{56.7}{14.4}
  \plot{56.1}{15.6}
  \plot{53.7}{20.4}
  \plot{50.1}{27.6}
  \plot{56.1}{16.}
  \plot{54.9}{18.4}
  \plot{54.9}{18.8}
  \plot{50.1}{28.4}
  \plot{42.9}{42.8}
  \plot{57.3}{14.8}
  \plot{56.1}{17.2}
  \plot{52.5}{24.4}
  \plot{57.3}{15.2}
  \plot{55.5}{18.8}
  \plot{53.7}{22.4}
  \plot{48.9}{32.}
  \plot{57.3}{15.6}
  \point{54.9}{20.4}
  \point{50.1}{30.}
  \plot{57.3}{16.}
  \plot{56.1}{18.4}
  \plot{53.7}{23.2}
  \plot{57.7}{15.6}
  \plot{57.3}{16.4}
  \plot{56.1}{18.8}
  \plot{53.7}{23.6}
  \plot{54.9}{21.6}
  \plot{57.3}{17.2}
  \plot{54.1}{23.6}
  \plot{50.1}{31.6}
  \plot{57.3}{17.6}
  \plot{53.7}{24.8}
  \plot{51.3}{29.6}
  \plot{57.3}{18.}
  \plot{52.5}{27.6}
  \plot{56.1}{20.8}
  \plot{57.3}{18.8}
  \plot{52.5}{28.4}
  \plot{50.1}{33.2}
  \plot{58.5}{16.8}
  \plot{56.1}{21.6}
  \plot{58.5}{17.2}
  \plot{57.3}{19.6}
  \plot{56.1}{22.}
  \plot{54.9}{24.4}
  \plot{58.5}{17.6}
  \plot{57.9}{18.8}
  \plot{56.1}{22.4}
  \plot{51.3}{32.}
  \plot{54.9}{25.2}
  \plot{46.5}{42.4}
  \plot{58.5}{18.8}
  \plot{53.7}{28.4}
  \plot{50.1}{35.6}
  \plot{59.1}{18.}
  \plot{58.5}{19.2}
  \point{57.3}{21.6}
  \plot{58.5}{19.6}
  \plot{58.1}{20.4}
  \plot{57.3}{22.}
  \plot{56.1}{24.4}
  \point{54.9}{26.8}
  \plot{54.9}{27.6}
  \plot{59.1}{19.6}
  \plot{33.3}{71.6}
  \plot{58.5}{21.6}
  \plot{56.7}{25.2}
  \point{58.5}{22.}
  \plot{57.3}{24.4}
  \plot{50.1}{38.8}
  \plot{58.5}{22.4}
  \plot{57.3}{24.8}
  \plot{59.7}{20.4}
  \plot{56.1}{27.6}
  \plot{58.5}{23.2}
  \plot{60.1}{20.4}
  \plot{59.7}{21.2}
  \point{59.1}{22.4}
  \plot{58.5}{23.6}
  \point{54.9}{30.8}
  \plot{45.3}{50.}
  \plot{58.5}{24.}
  \plot{60.3}{20.8}
  \plot{59.7}{22.}
  \plot{57.3}{26.8}
  \plot{59.7}{22.4}
  \plot{58.5}{24.8}
  \plot{60.3}{21.6}
  \plot{57.3}{27.6}
  \plot{60.5}{21.6}
  \plot{59.7}{23.2}
  \plot{56.1}{30.4}
  \plot{59.7}{23.6}
  \plot{50.1}{42.8}
  \plot{59.7}{24.4}
  \plot{60.9}{22.4}
  \plot{60.3}{24.}
  \plot{60.9}{23.2}
  \plot{59.7}{25.6}
  \plot{60.9}{24.}
  \plot{60.9}{24.8}
  \plot{60.9}{25.2}
  \plot{59.7}{27.6}
  \plot{58.5}{30.4}
  \plot{50.1}{47.6}
  \plot{45.3}{57.2}
  \plot{61.5}{25.2}
  \plot{60.9}{26.8}
  \plot{62.1}{24.8}
  \plot{53.7}{42.4}
  \plot{56.1}{38.}
  \plot{61.5}{27.6}
  \plot{60.9}{28.8}
  \plot{62.1}{26.8}
  \plot{62.7}{26.}
  \plot{56.1}{39.2}
  \plot{62.1}{27.6}
  \plot{60.9}{30.4}
  \plot{62.1}{28.4}
  \plot{57.3}{38.}
  \plot{58.5}{36.8}
  \plot{38.1}{78.8}
  \plot{60.9}{34.8}
  \plot{63.7}{30.}
  \plot{62.1}{33.2}
  \plot{57.3}{42.8}
  \plot{50.1}{57.2}
  \plot{60.9}{36.}
  \plot{64.5}{29.2}
  \plot{63.3}{32.4}
  \plot{64.5}{30.4}
  \plot{64.9}{30.}
  \plot{54.9}{50.}
  \plot{60.9}{38.4}
  \plot{64.5}{31.6}
  \plot{64.5}{32.4}
  \plot{64.5}{33.2}
  \plot{63.9}{34.4}
  \plot{64.5}{34.}
  \plot{63.3}{36.8}
  \plot{63.3}{37.6}
  \plot{65.7}{33.2}
  \plot{65.7}{34.4}
  \plot{66.3}{33.6}
  \plot{54.9}{57.2}
  \plot{65.7}{38.}
  \plot{62.1}{45.2}
  \plot{60.9}{48.}
  \plot{65.7}{39.2}
  \plot{66.9}{37.6}
  \point{64.5}{42.8}
  \plot{67.5}{37.2}
  \plot{63.3}{45.6}
  \plot{66.9}{39.6}
  \plot{68.1}{38.}
  \plot{65.7}{44.8}
  \plot{68.1}{45.6}
  \plot{70.5}{42.4}
  \plot{70.5}{44.8}
  \plot{69.3}{50.}
  \plot{70.5}{48.}
  \plot{71.7}{47.6}
  \plot{72.9}{51.2}
  \plot{74.1}{49.2}
  \plot{50.1}{100.4}
  \plot{62.1}{78.8}
  \plot{74.1}{57.2}
  \plot{66.9}{71.6}
  \plot{76.5}{55.6}
  \plot{77.1}{54.8}
  \plot{74.1}{66.8}
  \plot{81.3}{71.6}
  \plot{86.1}{78.8}
  \plot{98.1}{100.4}
  \mirrorline{50.1}{1.5}{103.5}
  \hskip 100 mm \hfill $$
\smno
{\pano \figindents \hskip 0.5cm Figure 1 \hskip 0.5cm mirror plot \pano}
\pano
Additionally, the GSO projection admits even more redundancies than (ii) and
(iii) like the well-known identities
$$10_E\cong1_A\otimes2_A,\quad 28_E\cong1_A\otimes3_A,\quad
          4_D\cong1_A\otimes1_A\eqno(4.2)$$
which can be easily checked in the Landau-Ginzburg formulation [\schwarz].
These identifications are valid for the diagonal $m$-invariant
$k+2=1\cdot(k+2)$ and for its mirror $k+2=(k+2)\cdot1$. Further
identifications can be read off from Table 1 whenever different tensor
products produce equal massless spectra:
$$10_{E,3\cdot4}\cong1_{A,3\cdot1}\otimes2_{A,1\cdot4},\quad
  4_{D,2\cdot3}\cong1_{A,1\cdot3}\otimes1_{A,1\cdot3}\cong4_{D,1\cdot6}
  \eqno(4.3)$$
and the corresponding mirror relations. For $28_E$ similar identities will
hold.
\meno
\section{5.\ Summary}
\pano
In this note we have investigated to which extent one can use $m$-invariants
for the construction of $N=1$ space-time string vacua. It turned out that
some conditions derived from string theory limited the number of models
drastically. We gave an example of a class of modular invariant $N=2$
supersymmetric partition functions which failed to allow  the spectral
flow automorphism  to map the $NS$ sector onto the $R$ sector. For the
surviving models we calculated the massless string spectrum showing
that almost half of all orbifold models can also be obtained using
$m$-invariants. Furthermore, this subclass of theories exhibits exact mirror
symmetry.
\smno
{\tt Acknowledgements:}\ It is a pleasure to thank L.\ Dolan, W.\ Nahm
and R.\ Schimmrigk for discussion. This work is supported by U.S.\ DOE
grant No.\ DE-FG05-85ER-40219.
\meno
\section{Appendix}
\pano
The 76 orbifold spectra are listed in Table 1. $n_{27}$ denotes the number of
generations, $n_{\o{27}}$ of antigenerations, $n_1$ of singlets and $n_g$
of gauge bosons. $\chi=2(n_{\o{27}}-n_{27})$ means the Euler number of the
underlying Calabi-Yau manifold. The $m$-invariants for a model
$k_1,\ldots,k_r$ are labeled by divisors of $k_i+2$ for each $i$. For brevity,
the spectra with $\chi>0$ are omitted. These missing spectra can be obtained
by the flip (4.1).
\bigno
\pano
\cl{\vbox{
\hbox{\vbox{\offinterlineskip
\def\tablespace
 {height2pt&\omit&&\omit&&\omit&&\omit&&\omit&&\omit&&\omit&&\omit&\cr}
\def\tablerule{\tablespace\noalign{\hrule height1pt}\tablespace}
\hrule\halign{&\vrule#&\strut\hskip0.1cm\hfil#\hfill\hskip0.1cm\cr
\tablespace
& $n_{27}$ && $n_{\o{27}}$ && $n_1$ && $n_g$ && $\chi$ && tensor product
 && $l$-invariant && $m$-invariant &\cr
\tablerule
& 37 && 7 && 200 && 4 && $-60$ && 1 4 4 4 4 && A A A A A && 1 2 2 2 2 & \cr
\tablespace
& 67 && 7 && 267 && 3 && $-120$ && 3 8 8 8 && A A A A && 1 2 2 2 & \cr
\tablespace
& 27 && 9 && 193 && 3 && $-36$ && 2 10 10 10 && A D D D && 1 3 3 3 & \cr
\tablespace
& 63 && 9 && 265 && 3 && $-108$ && 2 10 10 10 && A A A A && 1 3 3 3 & \cr
\tablespace
& 40 && 10 && 219 && 3 && $-60$ && 4 4 10 10 && A A A D && 2 2 4 4 & \cr
\tablespace
& 23 && 11 && 212 && 4 && $-24$ && 1 2 2 10 10 && A A A D D && 1 4 4 4 4 & \cr
\tablespace
& && && && && && 2 10 10 10 && A E D D && 4 4 4 4 & \cr
\tablespace
& 47 && 11 && 244 && 4 && $-72$ && 1 2 2 10 10 && A A A A A && 3 1 1 3 3 & \cr
\tablespace
& && && && && && 2 10 10 10 && A E A A && 1 3 3 3 & \cr
\tablespace
& 101 && 11 && 401 && 3 && $-180$ && 1 16 16 16 && A A A A && 1 2 2 2 & \cr
\tablespace
& 30 && 12 && 215 && 5 && $-36$ && 2 2 4 4 4 && A A A A A && 1 1 3 3 3 & \cr
\tablespace
& 25 && 13 && 188 && 4 && $-24$ && 1 2 2 10 10 && A A A D A && 3 1 1 3 3 & \cr
}}}
}}
\pano
\cl{\vbox{
\hbox{\vbox{\offinterlineskip
\def\tablespace
 {height2pt&\omit&&\omit&&\omit&&\omit&&\omit&&\omit&&\omit&&\omit&\cr}
\halign{&\vrule#&\strut\hskip0.1cm\hfil#\hfill\hskip0.1cm\cr
& && && && && && 2 10 10 10 && A E D A && 1 3 3 3 & \cr
\tablespace
& 97 && 13 && 405 && 3 && $-168$ && 1 10 16 34 && A D A A, A A A D
                 && 1 4 2 4 & \cr
\tablespace
& 27 && 15 && 212 && 4 && $-24$ && 1 2 4 4 10 && A A A A A, A A A A D
                 && 1 4 2 2 4 & \cr
\tablespace
& && && && && && 4 4 10 10 && A A E A, A A E D && 2 2 4 4 & \cr
\tablespace
& 23 && 17 && 205 && 3 && $-12$ && 2 10 10 10 && A D D A && 1 3 3 3 & \cr
\tablespace
& 35 && 17 && 229 && 3 && $-36$ && 2 10 10 10 && A D A A && 1 3 3 3 & \cr
\tablespace
& 36 && 18 && 223 && 3 && $-36$ && 4 4 10 10 && A A A A, A A D D
                && 2 2 4 4 & \cr
\tablespace
& 90 && 18 && 415 && 3 && $-144$ && 1 8 18 58 && A A D A, A A A D
                && 1 2 4 4, 3 2 4 12 & \cr
\tablespace
& 19 && 19 && 204 && 6 && 0 && 1 1 2 2 4 4 && A A A A A A
     && 1 1 4 4 2 2, 3 3 1 1 3 3 & \cr
\tablespace
& && && && && && 1 2 4 4 10 && A A A A E && 1 4 2 2 4, 3 1 3 3 3 & \cr
\tablespace
& && && && && && 2 2 4 4 4 && A A D A A
  && 1 1 3 3 3, 1 1 6 3 3 & \cr
\tablespace
& && && && && && && && 4 4 1 2 2, 4 4 2 2 2 & \cr
\tablespace
& && && && && && 4 4 10 10 && A A E E && 2 2 4 4, 3 3 3 3 & \cr
\tablespace
& 46 && 22 && 309 && 3 && $-48$ && 2 4 16 34 && A A D A, A A D D
                && 4 2 1 4, 4 2 2 4 & \cr
\tablespace
& 29 && 23 && 223 && 3 && $-12$ && 4 4 8 13 && A A A A
                && 2 2 2 1, 2 2 10 5 & \cr
\tablespace
& 30 && 24 && 247 && 3 && $-12$ && 2 4 16 34 && A A E A, A A E D
                && 4 2 1 4, 4 2 2 4 & \cr
\tablespace
& 60 && 24 && 325 && 3 && $-72$ && 2 4 22 22 && A A A A, A A D D
                && 1 3 3 3 & \cr
\tablespace
& 25 && 25 && 221 && 3 && 0 && 4 4 6 22 && A A D A, A A A D
   && 2 2 8 8, 3 3 1 3 & \cr
\tablespace
& 39 && 27 && 277 && 3 && $-24$ && 3 4 8 28 && A A A A
               && 1 2 2 2, 1 6 2 6 & \cr
\tablespace
& 75 && 27 && 425 && 3 && $-96$ && 1 10 14 46 && A A A A, A A D D
               && 3 3 1 3 & \cr
\tablespace
& 34 && 28 && 257 && 3 && $-12$ && 2 8 8 18 && A A A A, A A A D
               && 4 2 2 4 & \cr
\tablespace
& 65 && 29 && 355 && 3 && $-72$ && 1 10 16 34 && A A A A, A D A D
               && 1 4 2 4 & \cr
\tablespace
& 60 && 30 && 345 && 3 && $-60$ && 1 12 12 40 && A A A A
               && 1 2 2 2, 3 2 2 6 & \cr
\tablespace
& 38 && 32 && 299 && 3 && $-12$ && 4 4 5 40 && A A A A
               && 2 2 1 2, 2 2 7 14 & \cr
\tablespace
& 44 && 32 && 301 && 3 && $-24$ && 1 13 13 28 && A A A A, A A A D
               && 1 5 5 5, 1 5 5 10 & \cr
\tablespace
& 47 && 35 && 337 && 3 && $-24$ && 1 10 18 28 && A A A A, A D D A
       && 3 3 1 3, 3 3 5 15 &\cr
\tablespace
& 179 && 35 && 791 && 3 && $-288$ && 1 5 82 82 && A A A A, A A D D
         && 3 1 3 3 & \cr
\tablespace
& && && && && && && A A D A && 3 1 12 12 & \cr
\tablespace
& 61 && 37 && 413 && 3 && $-48$ && 2 4 13 58 && A A A A, A A A D
        && 4 2 1 4, 4 2 5 20 & \cr
\tablespace
& 44 && 38 && 333 && 3 && $-12$ && 2 4 16 34 && A A A A, A A A D
        && 1 3 9 9 & \cr
\tablespace
& 62 && 38 && 377 && 3 && $-48$ && 1 8 18 58 && A A A A, A A D D
             && 1 2 4 4, 3 2 4 12 & \cr
\tablespace
& 91 && 43 && 505 && 3 && $-96$ && 1 6 34 70 && A A A A, A D A D
             && 3 1 9 9 & \cr
\tablespace
& 63 && 51 && 457 && 3 && $-24$ && 1 10 12 82 && A A A A, A D A D
             && 3 3 1 3, 3 3 7 21 & \cr
\tablespace
& 101 && 53 && 565 && 3 && $-96$ && 1 6 28 118 && A D A A, A A A D
      && 1 8 2 8, 1 8 10 40 & \cr
\tablespace
& 61 && 55 && 447 && 3 && $-12$ && 1 8 16 88 && A A A A
             && 3 1 9 9, 3 5 9 45 & \cr
\tablespace
& 63 && 63 && 473 && 3 && 0 && 1 6 34 70 && A D A A, A A A D
             && 1 8 4 8, 3 1 9 9 & \cr
\tablespace
& 75 && 75 && 565 && 3 && 0 && 1 6 28 118 && A A A A, A D A D
                 && 1 8 2 8, 3 1 15 15 & \cr
\tablespace}\hrule}}
\smno
\hbox{\hskip 0.5cm Table 1 \hskip 0.5cm non-Gepner spectra}}}
\pano
\vfill
\eject
\section{References}
\pano
\bibitem{\berg} P.\ Berglund and T.\ H\"ubsch, {\it A generalized construction
of mirror manifolds}, Nucl.\ Phys.\ {\bf B393} (1993) 377
\bibitem{\banks} T.\ Banks, L.J.\ Dixon, D.\ Friedan and E.\ Martinec,
{\it Phenomenology and conformal field theory, or Can string theory predict
the weak mixing angle}?, Nucl.\ Phys.\ {\bf B299} (1988) 613
\bibitem{\selfa} R.\ Blumenhagen, {\it $N=2$ supersymmetric \w-algebras},
Nucl.\ Phys.\ {\bf B405} (1993) 744
\bibitem{\selfb} R.\ Blumenhagen and R.\ H\"ubel, {\it A note on
representations of $N=2$ \sw-algebras},
Mod.\ Phys.\ Lett.\ {\bf A9} (1994) 3193
\bibitem{\selfc} R.\ Blumenhagen and A.\ Wi{\ss}kirchen, {\it Generalized
string functions of $N=1$ space-time supersymmetric string vacua}, preprint
BONN-TH-95-03, IFP-503-UNC, hep-th/9501072,
to be published in Phys.\ Lett.\ {\bf B}
\bibitem{\cande} P.\ Candelas, G.T.\ Horowitz, A.\ Strominger and E.\ Witten,
{\it Vacuum configurations for superstrings},
Nucl.\ Phys.\ {\bf B258} (1985) 46
\bibitem{\candz} P.\ Candelas, M.\ Lynker and R.\ Schimmrigk, {\it Calabi-Yau
manifolds in weighted $P(4)$}, Nucl.\ Phys.\ {\bf B341} (1990) 383
\bibitem{\ciz} A.\ Cappelli, C.\ Itzykson and J.P.\ Zuber, {\it The $ADE$
classification of minimal and $A^{(1)}_1$ conformal invariant theories},
Commun.\ Math.\ Phys.\ {\bf 113} (1987) 1
\bibitem{\egu} T.\ Eguchi, H.\ Ooguri, A.\ Taormina and S.K.\ Yang,
{\it Superconformal algebras and string compactification on manifolds
with $SU(n)$ holonomy}, Nucl.\ Phys.\ {\bf B315} (1989) 193
\bibitem{\fonti} A.\ Font, L.E.\ Ib\'a$\tilde{\rm n}$ez, F.\ Quevedo and
A.\ Sierra, {\it Twisted $N=2$ coset models:\ discrete torsion and asymmetric
heterotic string compactifications}, Nucl.\ Phys.\ {\bf B337} (1990) 119
\bibitem{\fuchse} J.\ Fuchs, A.\ Klemm, C.M.A.\ Scheich and M.G.\ Schmidt,
{\it Spectra and symmetries of Gepner models compared to Calabi-Yau
compactifications}, Ann.\ Phys.\ {\bf 204} (1990) 1
\bibitem{\fuchsz} J.\ Fuchs, A.\ Klemm and M.G.\ Schmidt, {\it Orbifolds by
cyclic permutation in Gepner type superstring and in the corresponding
Calabi-Yau manifolds}, Ann.\ Phys.\ {\bf 214} (1992) 231
\bibitem{\gep} D.\ Gepner, {\it Space-time supersymmetry in compactified
string theory and superconformal models}, Nucl.\ Phys.\ {\bf B296} (1988) 757
\bibitem{\gepq} D.\ Gepner and Z.\ Qiu, {\it Modular invariant partition
functions for parafermionic theories}, Nucl.\ Phys.\ {\bf B285} (1987) 423
\bibitem{\gre} B.R.\ Greene and M.R.\ Plesser, {\it Duality in Calabi-Yau
moduli space}, Nucl.\ Phys.\ {\bf B338} (1990) 15
\bibitem{\int} K.\ Intriligator and C.\ Vafa, {\it Landau-Ginzburg orbifolds},
Nucl.\ Phys.\ {\bf B339} (1990) 95
\bibitem{\klsch} A.\ Klemm and R.\ Schimmrigk, {\it Landau-Ginzburg string
vacua}, Nucl.\ Phys.\ {\bf B411} (1994) 559
\bibitem{\kreuzm} M.\ Kreuzer and H.\ Skarke, {\it No mirror symmetry
in Landau-Ginzburg spectra}!, Nucl.\ Phys.\ {\bf B388} (1992) 113
\bibitem{\kreuza} M.\ Kreuzer and H.\ Skarke, {\it $ADE$ string vacua with
discrete torsion}, Phys.\ Lett.\ {\bf B318} (1993) 305
\bibitem{\kreuzs} M.\ Kreuzer and A.N.\ Schellekens, {\it Simple currents
versus orbifolds with discrete torsion -- a complete classification},
Nucl.\  Phys.\ {\bf B411} (1994) 97
\bibitem{\kreuzl} M.\ Kreuzer and H.\ Skarke, {\it Landau-Ginzburg orbifolds
with discrete torsion}, preprint ITP-UH-16/94, TUW-94/20, hep-th/9412033
\bibitem{\lerche} W.\ Lerche and N.P.\ Warner, {\it Index theorems in $N=2$
superconformal theories}, Phys.\ Lett.\ {\bf B205} (1988) 471
\bibitem{\lut} C.A.\ L\"utken and G.G.\ Ross, {\it Taxonomy of heterotic
superconformal field theories}, Phys.\ Lett.\ {\bf B213} (1988) 152
\bibitem{\lyne} M.\ Lynker and R.\ Schimmrigk, {\it On the spectrum of
$\,(2,2)$ compactifications of the heterotic string on conformal field
theories}, Phys.\ Lett.\ {\bf B215} (1988) 681
\bibitem{\lynz} M.\ Lynker and R.\ Schimmrigk, {\it $ADE$ quantum Calabi-Yau
manifolds}, Nucl.\ Phys.\ {\bf B339} (1990) 121
\bibitem{\lynd} M.\ Lynker and R.\ Schimmrigk, {\it Landau-Ginzburg theories
as orbifolds}, Phys.\ Lett.\ {\bf B249} (1990) 237
\bibitem{\mart} E.\ Martinec, {\it Algebraic geometry and effective
Lagrangians}, Phys.\ Lett.\ {\bf B217} (1989) 431
\bibitem{\qiu} Z.\ Qiu, {\it Modular invariant partition functions for $N=2$
superconformal field theories}, Phys.\ Lett.\ {\bf B198} (1987) 497
\bibitem{\sche} A.N.\ Schellekens and S.\ Yankielowicz, {\it Extended chiral
algebras and modular invariant partition functions},
Nucl.\ Phys.\ {\bf B327} (1989) 673
\bibitem{\schz} A.N.\ Schellekens and S.\ Yankielowicz, {\it New modular
invariants for $N=2$ tensor products and four-dimensional strings},
Nucl.\ Phys.\ {\bf B330} (1990) 103
\bibitem{\scht} A.N.\ Schellekens and S.\ Yankielowicz,
CERN-TH.5440S/89 + NIKHEF-H/5440T/89 (unpublished table supplements)
\bibitem{\schwarz} J.H.\ Schwarz, {\it Superconformal symmetry and
superstring compactification}, Int.\ Jour.\ Mod.\ Phys.\ {\bf A4} (1989) 2653
\bibitem{\schwim} A.\ Schwimmer and N.\ Seiberg, {\it Comments on the
$N=2,3,4$ superconformal algebras in two dimensions},
Phys.\ Lett.\ {\bf B184} (1987) 191
\bibitem{\vw} C.\ Vafa and N.P.\ Warner, {\it Catastrophes and the
classification of conformal theories}, Phys.\ Lett.\ {\bf B218} (1989) 51
\bibitem{\zog} P.\ Zoglin, {\it Heterotic string compactifications using
minimal $N=2$ superconformal field theory quotient models},
Phys.\ Lett.\ {\bf B218} (1989) 444
\vfill
\end